\documentstyle[preprint,revtex,eqsecnum]{aps}
\begin{document}

\draft

\begin{title}
Primordial magnetic fields from pseudo-Goldstone bosons
\end{title}
\author{W.~Daniel~Garretson, George~B.~Field and Sean~M.~Carroll}
\begin{instit}
Harvard-Smithsonian Center for Astrophysics \\
60 Garden Street, Cambridge, Massachussetts 02138 USA
\end{instit}
\receipt{21 July 1992}

\begin{abstract}
The existence of large-scale magnetic fields in galaxies is well
established, but there is no accepted mechanism for generating a
primordial field which could grow into what is observed today.  We
discuss a model which attempts to account for the necessary primordial
field by invoking a pseudo-Goldstone boson coupled to electromagnetism.
The evolution of this boson during inflation generates a magnetic
field; however, it seems difficult on rather general grounds to obtain
fields of sufficient strength on astrophysically interesting scales.
\end{abstract}

\pacs{98.80.Cq, 98.60.Jk, 95.30.Cq, 14.80.Gt}

\narrowtext

\section{INTRODUCTION}

The existence of a magnetic field of $\sim 10^{-6}$ gauss in
our galaxy and other galaxies is well established \cite{zrs}.
The explanation of how such a
magnetic field arose is, however, far from certain.  While the
creation and evolution of stellar magnetic fields is fairly well
understood, the extension of these theories to
galaxies suffers from problems relating to both scale length and
time scales.

Zeldovich, Ruzmaikin, and Sokoloff \cite{zrs} and
Parker \cite{parker} discuss the
origin and effects of magnetic fields in
the universe and draw the conclusion that the galactic field arises
from a dynamo mechanism.  The dynamo model requires a seed
field at the epoch of galaxy formation which is coherent over a scale
of $\sim 1$ Mpc.  We can parameterize the strength of a primordial
field by $r~=~\rho_B/\rho_\gamma$, the ratio
of the energy density $\rho_B~=~B^2/8\pi$ in the
magnetic field relative to that of the background radiation
$\rho_\gamma$.  (This ratio is constant while the universe is a good
conductor, which is almost always \cite{tw}.)
Then the field required to seed a galactic dynamo satisfies
$r\agt 10^{-31}$, corresponding
to an intergalactic field at the epoch of galaxy formation ($z \sim$
3--5) of $\agt 10^{-20}$ gauss.
The implications of this requirement in terms of the origin
of such fields and their possible effect on star formation (or the
early history of the galaxy in general) are discussed by Rees
\cite{rees}.

Kulsrud \cite{kulsrud} has argued that the galactic dynamo
explanation is fundamentally flawed in that
the mean square deviations of the
magnetic field will grow much faster than the mean field itself,
resulting in a disordered field with a much smaller mean field
strength than would naively be expected.  Kulsrud then
discusses the possibility that the magnetic field originated in the
early universe and was embedded in the medium out of which the
galaxies ultimately formed. In this case, Kulsrud argues that the
intergalactic field would need to be $\agt
 10^{-12}$ gauss at the epoch of galaxy formation (giving
$r\agt0^{-15}$)
in order to account for the observed interstellar field.

Any dynamo theory requires a mechanism for generating the required
seed field; however, compelling mechanisms have been elusive.
The hot plasma in the early universe is
highly conducting and thus should strongly inhibit the growth of
a magnetic field, even to
$r~\sim~10^{-31}$. Furthermore, any hypothetical process at work in
the very early universe must be able to produce fields with
characteristic length scales much larger then the horizon at that
time, in order to correspond to galactic scales today.

The advent of inflation \cite{inflation} has opened the
door to new possibilities for generating a primordial magnetic
field.  There are two key features of an inflationary
universe that make the possibility of creating a
magnetic field during this time particularly attractive. First, if
there were an inflationary epoch at very early times in the universe,
the exponential expansion would have reduced the conductivity to a
negligible value by reducing the charged particle density, thus
allowing the creation of a substantial magnetic
field.  If this field were then frozen into the plasma created
during the subsequent reheating of the universe, it will be
supported by the effectively infinite conductivity of the
plasma so that its strength will decay only as inverse square of the
scale factor.

Second, if inflation did occur, then the entire observable universe
today was, at some point in the early universe, contained entirely
within the particle horizon.  It would then be possible to use
physical mechanisms operating on a scale smaller than the horizon
to generate magnetic fields that are coherent over macroscopic scales
today, an opportunity which is not available in models of the
early universe without inflation.

Turner and Widrow \cite{tw} (TW) investigated the possibility
that quantum
fluctuations during an inflationary epoch might have generated a
magnetic field that could be sustained after the
wavelength of interest crossed beyond the horizon and thus give the
observed field today.  TW considered coupling the electromagnetic
field to the curvature tensor so as to
amplify the fluctuation-induced field, but found satisfactory
results could be obtained only at the expense of breaking gauge
invariance.

Ratra \cite{ratra} has argued that it is possible
to generate a magnetic field
with present field strength of
$\agt 10^{-10}$ gauss on a scale
of 1/1000 the present Hubble radius by coupling a scalar field $\Phi$
to the electromagnetic potential $A_\mu$ through a term of the form
$e^{\Phi}F_{\mu\nu}F^{\mu\nu}$, where
$F_{\mu\nu} = \partial_\mu A_\nu - \partial_\nu A_\mu$ is the
electromagnetic field strength.  This
would be sufficient to explain the galactic magnetic field,
but it remains to be seen whether or not this coupling could arise
naturally in realistic particle physics models.

TW also suggested that the magnetic field could be sustained by
coupling the EM field strength to a pseudoscalar axion field
$\phi$ via an interaction term in the Lagrange density of the form
\begin{equation}
  {\cal L} = {g_{\phi\gamma\gamma} \over 4}\phi
  F_{\mu\nu}{\widetilde F}^{\mu\nu} \ ,
  \eqnum{1.1}
\end{equation}
where ${\widetilde F}^{\mu\nu}\equiv {1\over 2} \epsilon^{\mu\nu
\rho\sigma}F_{\rho\sigma}$ is the dual of $F_{\mu\nu}$, and
$g_{\phi\gamma\gamma} = (\alpha/2\pi)/f$, where $f$ is a coupling
constant with units of mass
and $\alpha$ is the fine structure constant.  However, they did not
complete the necessary analysis to show whether or not this might
indeed be the case.

An interaction of this form has been studied by Carroll and Field
\cite{cf} (see also \cite{cfj}), who
found that the evolution of a Fourier mode of the magnetic field with
wavenumber $k$ is governed by
\begin{equation}
  {d^2F_\pm \over d\eta^2} + \left( k^2 \pm g_{\phi\gamma\gamma} k
  {d\phi \over d\eta} \right)F_{\pm} = 0 \ ,\eqnum{1.2}
\end{equation}
where $F_\pm = a^2(B_y \pm iB_z)$ (the $\pm$ refers to different
circular polarization modes of the magnetic field), $d\eta = dt/a$,
and $a$ is the scale factor of the universe (normalized so that $a_0 =
1$ where $a_0$ is the value of $a$ today).  One (or both) of the
polarization modes will be unstable for $k < g_{\phi\gamma\gamma}
|d\phi/d\eta|$, where both polarization modes
can be unstable to exponential growth if $\phi$ is oscillatory.
Thus, if such a scalar field exists during inflation (perhaps the
inflaton itself \cite{ffo}) with the above coupling,
this might provide a mechanism for generating a
substantial magnetic field.

Here we will consider a generalization of the possibility
suggested by TW \cite{tw}, coupling the photon
to an arbitrary pseudo-Goldstone boson
(PGB) rather than the axion of QCD.  The PGB is characterized by
a spontaneous symmetry breaking scale $f$ (as above) and a soft
explicit symmetry breaking scale $\Lambda$ (see \hbox{Sec.~II}).  We
find that significant growth occurs only at a temperature near
$\Lambda$, and that the
magnetic field strength thus generated cannot give an astrophysically
interesting field at the end of inflation.

We should point out that, while our notation throughout this paper
suggests that we are working with the photon of the standard $U(1)_{em}$
symmetry, the photon as a separate $U(1)$ gauge boson will not exist
at high temperature, since the $SU(2)\times U(1)$ gauge
symmetry will not have been broken.  Nevertheless, our results should
be correct up to factors of order unity simply because the $U(1)$
hypercharge symmetry projects onto the photon with a multiplicative
factor of $\cos\theta_w \approx 0.88$ at the electroweak phase
transition.

We will use units in which $\hbar = c = k_B = 1$, such that $G =
m_{pl}^{-2}$, where $m_{pl} \approx 1.22 \times 10^{19}$ GeV is the
Planck mass.

\section{THE SET-UP}

In this section we briefly review the essentials of inflation, as well
as the physics of pseudo-Goldstone bosons and their couplings.

Throughout this paper, we will assume that the universe is in a
spatially flat FRW cosmology in which the metric is given by
\begin{equation}
  ds^2 = a^2(\eta)(-d\eta^2 + d{\bf x}^2) \ ,\eqnum{2.1}
\end{equation}
where {\bf x} represents the standard Cartesian 3-space
(comoving) coordinates.
In addition, we will assume that the universe is a perfect fluid with
the equation of state $p = \gamma\rho$, where $p$ is the pressure,
$\rho$ is the total energy density, and $\gamma$ is a constant.
Using this equation and
energy-momentum conservation, it is straightforward to show that $\rho
\propto a^{-3(1+\gamma)}$ which, from Einstein's equation, gives $a
\propto t^{2/3(1+\gamma)}$.  In order to explain the horizon problem,
we require that, at some time in the past, the scale factor was
growing faster than the horizon ($H^{-1}$, where $H = {\dot a}/a$ is
the Hubble parameter and an overdot denotes differentiation with
respect to physical time $t$).  Thus,
we require $-1 \le \gamma < -{1 \over 3}$.  For simplicity in the
following discussion we will restrict ourselves to inflation in which
$\gamma = -1$.  This value for $\gamma$ gives the
best possible conditions for generating a magnetic field, since
the amount of inflation from the time that a given comoving wavelength
crosses outside the horizon is minimal in this case, so
this does not limit the validity of our results.

At the end of inflation, the universe enters a reheating phase, in
which the energy density is matter-dominated.  As a simplifying
assumption, we take the process of reheating to be instantaneous, such
that the universe goes directly from inflation to radiation
domination.  Once again this is a best-possible assumption, since the
magnetic field will decay more rapidly (relative to the total energy
density) during a matter-dominated phase, in which $\rho\propto
a^{-3}$.

Standard inflation is characterized by two parameters:
the mass scale
for the total energy density $M = \rho^{1/4}$ (note that this is a
constant since $\rho$ is constant during inflation with $\gamma = -1$);
and the temperature
$T_{RH}$ to which the universe reheats at the end of inflation.
$H$ is then given by
\begin{equation}
  H^2 = {4\pi\over 3m_{pl}^2}M^4 \ .\eqnum{2.2}
\end{equation}
If we assume that the universe expands adiabatically after
inflation so that the entropy per comoving volume element remains
constant, it can be shown (see e.g. \cite{kt}) that the total
expansion from the time a given comoving wavelength $\lambda$ (which
is equal to the physical wavelength today due to the normalization of
$a$) crosses outside the horizon until the end of inflation is given
by
\begin{equation}
  {a_{inf} \over a_\lambda} \simeq 10^{26} {\lambda \over {\rm
  Mpc}} \left( M^2 T_{RH} \over m_{pl}^3 \right)^{1 \over 3} \ ,
  \eqnum{2.3}
\end{equation}
where $a_{inf}$ is the value of $a$ at the end of inflation (but
before reheating), and $a_\lambda$ is the value of $a$ at the time
$\lambda$ crosses outside the horizon.  Expressing this in terms of
the number of $e$-foldings $N_\lambda$
($a_{inf}/a_\lambda = e^{N_\lambda}$) in the expansion, we have
\begin{equation}
  N_\lambda \simeq 48 + \ln\left(\lambda \over {\rm Mpc}\right) + {2
  \over 3}\ln\left( M \over 10^{14}\ {\rm GeV}\right) + {1 \over
  3}\ln\left( T_{RH} \over 10^{14}\ {\rm GeV}\right) \ . \eqnum{2.4}
\end{equation}
Our assumption that reheating lasts for a negligible time amounts to
setting $M=T_{RH}$.

In this paper we are concerned with the pseudo-Goldstone
boson $\phi$ of a spontaneously broken symmetry.  PGBs
are characterized by two mass scales: a large mass $f$ at which
the global symmetry from which the PGBs arise is spontaneously broken,
and a smaller scale $\Lambda$ at which the symmetry is explicitly
broken.  For concreteness, we imagine the breakdown of a
global $U(1)$ symmetry, resulting in the familiar Mexican hat --- the
radial degree of freedom gets a vacuum expectation value of order $f$,
and the angular degree of freedom becomes a massless boson $\phi$.
The hat is tilted by a small term of order $\Lambda$; the formerly
massless scalar $\phi$ becomes a PGB with mass of order
\begin{equation}
  m\approx\Lambda^2/f\ . \eqnum{2.5}
\end{equation}
Cosmological constraints on the parameters $f$ and $\Lambda$ have been
studied in \cite{fj}.

In many models, PGB's interact with fermions by coupling to the axial
vector current (for a review of PGB's and their couplings, see
\cite{kim}):
\begin{equation}
  {\cal L}_{\rm int}= {1\over 4f}\bar\psi\gamma^\mu
  \gamma_5\psi\partial_\mu\phi= {1\over 4f}J^\mu_5\partial_\mu\phi \ ;
  \eqnum{2.6}
\end{equation}
well-known examples include pions and axions.
Since the symmetry associated with the axial current (that is, chiral
symmetry) is anomalously
broken, the current itself is not conserved:
\begin{equation}
  \partial_\mu J^\mu_5={\alpha\over 2\pi}F_{\mu\nu}\widetilde
  F^{\mu\nu} \ ,\eqnum{2.7}
\end{equation}
where $\alpha$ is the fine structure constant.  Integrating by parts,
we find that the anomaly (2.7) induces a coupling
between $\phi$ and the electromagnetic field of the form (1.1).

A similar
situation occurs in the low-energy limit of string theory, which
involves an antisymmetric two-index tensor field $B_{\mu\nu}$
\cite{kim,gsw}.  The Lagrangian for
$B_{\mu\nu}$ includes a kinetic term
$H_{\mu\nu\rho}H^{\mu\nu\rho}$, where $H_{\mu\nu\rho}$ is an
antisymmetric field strength tensor.  The demand that the theory be
free of anomalies requires that the definition of $H_{\mu\nu\rho}$
include a term
involving gauge bosons (which we take to be abelian for simplicity):
\begin{equation}
  H_{\mu\nu\rho}=\partial_{[\mu}B_{\nu\rho]}-
  A_{[\mu}F_{\nu\rho]} \ ,\eqnum{2.8}
\end{equation}
where square brackets denote antisymmetrization.  In
four dimensions the equations of motion for $B_{\mu\nu}$ allow us to
recast its dynamics (at least semiclassically) in terms
of a pseudoscalar $\phi$ by the identification
\begin{equation}
  \partial_\mu\phi={1\over f}\epsilon_{\mu\alpha\beta\gamma}
  H^{\alpha\beta\gamma} \ .\eqnum{2.9}
\end{equation}
The interaction between this pseudoscalar and electromagnetism, as
implied by (2.8) and (2.9), can be described by an
effective Lagrangian of the form (1.1).  Thus, string theory offers
the possibility of PGB's of the type we discuss.

\section{BASIC REQUIREMENTS}

We envision a scenario in which the desired magnetic
field is entirely created
during inflation and then frozen into the post-reheat plasma,
so that the field strength decays as the inverse square of the scale
factor after reheating.  Thus, under the assumption that the dominant
component of background photons is created at reheating, both $\rho_B$
and $\rho_\gamma$ will decay as $a^{-4}$, such that
$r=\rho_B/\rho_\gamma$ is essentially constant after inflation.
As mentioned in the introduction, we seek $r = r_0
\agt 10^{-31}$ (to seed a
galactic dynamo) or $\agt
10^{-15}$ (to directly account for the
observed magnetic field) at scales of $\sim$~1~Mpc today.

For convenience we define $r$ during inflation as the ratio of
$\rho_B$ to the total (vacuum-dominated) energy density.
{}From above, we can readily calculate the required value of $r$ at
the time that the wavelength of interest (comoving scale 1 Mpc)
crosses outside the horizon in order to generate astrophysically
interesting fields today. This is given by
\begin{equation}
  r_{\rm Mpc} = \left( a_{inf} \over a_{\rm Mpc} \right)^4 r_0 \ ,
  \eqnum{3.1}
\end{equation}
or, using (2.3),
\begin{equation}
   r_{\rm Mpc} = 10^{104} \left( M \over m_{pl} \right)^4 r_0 \ .
  \eqnum{3.2}
\end{equation}
Since $r < 1$, we can use this to calculate the
maximum allowed value for $M$ such that there would be any hope for
meeting the above condition.  Using the values given above for $r_0$,
this gives $M \alt 10$ GeV for a galactic
dynamo, or $\alt 1$
MeV to directly account for the field.  It is important to realize
that these are extreme upper bounds that can only be obtained under
ideal situations (i.e. the magnetic field energy being comparable to
that in the vacuum at the time the wavelength of interest crosses the
horizon, and with negligible reheating).  Even so, the upper bound for
seeding the galactic dynamo is, at best, marginal (that is, while
current constraints on the time at which inflation might occur {\it
could} allow it to occur as late as a temperature of $\sim 1$ GeV
since we merely require the universe to be radiation dominated by the
epoch of primordial nucleosynthesis \cite{tw}, constraints arising
from baryogenesis will probably require
$M \agt 200$ GeV, at
least).  Thus, even if the magnetic field has an energy density
comparable to the background energy density when 1 Mpc crosses the
horizon during inflation, it will be too weak to explain the observed
magnetic field if its energy decays according to the normal $a^{-4}$.

In other words, in order to generate a significant magnetic field
during inflation, we require ``superadiabatic growth'' ---  that is,
a mechanism
that will continue to increase the energy density (or, at least,
decrease the rate of decay) of the magnetic field at a wavelength of 1
Mpc {\it after} 1 Mpc becomes super-horizon sized.  At first sight,
this may seem impossible due to the fact that such a mechanism must
apparently act in an acausal way.  However, it is possible that
inflation may create a field that is coherent over scales much larger
than the horizon, and that this field can subsequently generate a
magnetic field that is also coherent at super-horizon scales
simply by classical field interactions.

\section{EQUATIONS OF MOTION}

The Lagrange density for the photon and scalar field is
\begin{equation}
  {\cal L} = -\sqrt{g}\left( {1 \over 2} \nabla_\mu \phi \nabla^\mu
  \phi + V(\phi) + {1 \over 4}F_{\mu\nu}F^{\mu\nu} +
  {g_{\phi\gamma\gamma} \over 4}\phi
   F_{\mu\nu}{\widetilde F}^{\mu\nu}
  \right) \ , \eqnum{4.1}
\end{equation}
where $g = -{\rm det}(g_{\mu\nu})$ and $\nabla_\mu$
denotes the covariant derivative.  As
mentioned previously, we are considering only the $U(1)$ fields
and ignoring any possible effects from the nonabelian gauge fields.

The equations of motion for $\phi$ are
\begin{equation}
  -\nabla^\mu \nabla_\mu \phi + {dV(\phi) \over d\phi} =
  -{g_{\phi\gamma\gamma} \over 4} F_{\mu\nu}{\widetilde
  F}^{\mu\nu} \ ,\eqnum{4.2}
\end{equation}
and the equations of motion for $F_{\mu\nu}$ are
\begin{equation}
  \nabla_\mu F^{\mu\nu} = -g_{\phi\gamma\gamma}(\nabla_\mu
  \phi){\widetilde F}^{\mu\nu} \ ,\eqnum{4.3}
\end{equation}
along with the Bianchi identity
\begin{equation}
  \nabla_\mu {\widetilde F}^{\mu\nu} = 0 \ .\eqnum{4.4}
\end{equation}
The equations of motion are more transparent if we define the {\bf E}
and {\bf B} fields by
\begin{equation}
   F^{\mu\nu} = a^{-2}\pmatrix{ 0 & E_x & E_y & E_z \cr
	-E_x & 0 & B_z & -B_y \cr
	-E_y & -B_z & 0 & B_x \cr
	-E_z & B_y & -B_x & 0 \cr } \ .\eqnum{4.5}
\end{equation}
Then (4.2) becomes, after expanding the covariant derivatives,
\begin{equation}
  {\partial^2 \phi \over \partial \eta^2} + 2aH{\partial \phi
	\over \partial \eta} - \nabla^2 \phi + a^2
	{dV(\phi) \over d\phi} =
        g_{\phi\gamma\gamma} a^2{\bf E \cdot B} \ ,
  \eqnum{4.6}
\end{equation}
where $\nabla$ represents the usual 3-space gradient
(for comoving coordinates).  Similarly, (4.3) becomes
\begin{equation}
  {\partial \over \partial \eta}(a^2{\bf E}) - \nabla \times (a^2{\bf
   B}) = -g_{\phi\gamma\gamma}
  {\partial \phi \over \partial \eta}a^2{\bf B} -
  g_{\phi\gamma\gamma}(\nabla \phi) \times a^2{\bf E} \ , \eqnum{4.7}
\end{equation}
with
\begin{equation}
  \nabla \cdot {\bf E} = -g_{\phi\gamma\gamma}(\nabla \phi) \cdot
  {\bf B} \ ,\eqnum{4.8}
\end{equation}
while the Bianchi identity becomes
\begin{equation}
  {\partial \over \partial \eta}(a^2{\bf B}) + \nabla \times
  (a^2 {\bf E}) = 0 \ ,\eqnum{4.9}
\end{equation}
with
\begin{equation}
  \nabla \cdot {\bf B} = 0 \ . \eqnum{4.10}
\end{equation}

Since we are interested in the specific case where the background
space-time is inflating, we make the assumption that the spatial
derivatives of $\phi$ are negligible compared to the other terms (if
this is not the case at the beginning of inflation, any spatial
inhomogeneities will quickly be inflated away and this assumption will
quickly become very accurate).  Then, eliminating {\bf E} in the above
equations, we have
\begin{equation}
  \left({\partial^2 \over \partial \eta^2} - \nabla^2 -
  g_{\phi\gamma\gamma}{d\phi \over d \eta}\nabla \times \right)
  (a^2{\bf B}) = 0 \ .\eqnum{4.11}
\end{equation}
Taking the spatial fourier transform of this equation so that
\begin{equation}
  {\bf B}(\eta,{\bf k}) = {1 \over 2\pi}\int e^{i{\bf k \cdot x}}
  {\bf B}(\eta,{\bf x}) d^3{\bf x} \ ,\eqnum{4.12}
\end{equation}
and writing ${\bf F} = a^2 {\bf B}$, we then have
\begin{equation}
  {\partial^2 {\bf F} \over \partial \eta^2} + k^2{\bf F}
  - g_{\phi\gamma\gamma}{d \phi \over d \eta}i{\bf k}
  \times {\bf F} = 0 \ .\eqnum{4.13}
\end{equation}
Finally, if we take {\bf k} to point along the x-axis and define
$F_\pm = F_y \pm iF_z$, this becomes
\begin{equation}
   {\partial^2 F_\pm \over \partial \eta^2}
   + \left(k^2 \pm g_{\phi\gamma\gamma}{d \phi \over d
   \eta}k \right) F_\pm = 0 \ .\eqnum{4.14}
\end{equation}
We can similarly manipulate the equations in an attempt to produce an
expression for the evolution of {\bf E}.  It turns out, however, that
we cannot uncouple the {\bf E} field from the {\bf B} field.  We have,
in short,
\begin{equation}
  \left({\partial^2 \over \partial \eta^2} - \nabla^2 -
  g_{\phi\gamma\gamma}{d \phi \over d \eta}\nabla \times \right)
  (a^2{\bf E}) = -g_{\phi\gamma\gamma}{d^2\phi \over d\eta^2}a^2
  {\bf B}\ ,\eqnum{4.15}
\end{equation}
which, after taking the space fourier transform and defining ${\bf G}
= a^2{\bf E}$ and $G_\pm = G_y \pm iG_z$, becomes
\begin{equation}
  {\partial^2 G_\pm \over \partial \eta^2}
  + \left(k^2 \pm g_{\phi\gamma\gamma}{d \phi \over d
  \eta}k \right) G_\pm = -g_{\phi\gamma\gamma}
  {d^2\phi \over d\eta^2}F_\pm \ .
  \eqnum{4.16}
\end{equation}

In order to determine the evolution of {\bf E} and {\bf B}, we need to
know how $\phi$ evolves.  We
look first at the case where {\bf E} and {\bf B} make
a negligible contribution to the equation of motion for $\phi$.
Furthermore, we will consider the evolution after the time when the
explicit symmetry breaking for the PGB becomes important (at a
temperature scale $\Lambda$).  The potential for the angular degree of
freedom in a tilted Mexican hat is
\begin{equation}
  V(\phi) = \Lambda^4[1 - \cos(\phi/f)] \ .\eqnum{4.17}
\end{equation}
Since the details of the potential do not affect our results, we will
expand to lowest order: $V(\phi)\sim (\Lambda^4/2f^2)\phi^2$.
We again ignore spatial derivatives in $\phi$, as well as any
back reaction from the electromagnetic fields, to give
\begin{equation}
  {d^2\phi \over dt^2} + 3H{d\phi \over dt} + {\Lambda^4 \over
  f^2}\phi = 0 \ ,\eqnum{4.18}
\end{equation}
where we have used $d\eta = dt/a$ to write this in terms of physical
time rather than conformal time.  The general solution to (4.18) will
be approximately
\begin{eqnarray}
  \phi(t) &\approx & f\exp\left[-{1 \over 2}\left(3H \pm
  \sqrt{9H^2 - 4{\Lambda^4 \over f^2}}\right)(t - t_0)\right]
  \nonumber  \\
  & \propto & \cases{a^{-3/2}\sin\left({\Lambda^2 \over f}t\right)
  & for ${\Lambda^2 \over f} \gg H$, \cr
  \exp\left(-{\Lambda^4 \over 3Hf^2}t\right) & for ${\Lambda^2
  \over f} \ll H$, \cr} \eqnum{4.19}
\end{eqnarray}
where we have used the fact that $a \propto \exp{Ht}$.

Looking again at (4.14) we see that we will only have a growing
mode for the magnetic field if
\begin{equation}
  {d\phi \over d\eta} = a{d\phi \over dt} > k \ .\eqnum{4.20}
\end{equation}
Furthermore, since we want this growth to be as large as possible, we
will choose $k$ such that $k \sim g_{\phi\gamma\gamma} a{\dot
\phi}_{max}$ at the
time  of interest.  For our initial analysis, we will assume
$\Lambda^2/f \gg H$ so that $\phi$ is oscillating rapidly compared to
changes in $a$. Also, this implies that $a$ is essentially constant
over several oscillations in $\phi$, and thus we can write $\Delta\eta
\approx \Delta t/a$ where $a$ is constant for time intervals $\Delta t
\sim f/\Lambda^2$.

In order to estimate
the total growth in $F_\pm$, we note that, for a fraction $\epsilon$
(where $\epsilon$ is not necessarily small, although numerical
integration of these equations for some cases indicates that $\epsilon
\sim 0.1$)
of each period, we can write $F_\pm \propto e^{\alpha \Delta \eta}$
where $\alpha=\sqrt{g_{\phi\gamma\gamma} k d\phi/d\eta} \sim
ag_{\phi\gamma\gamma}\Lambda^2$ and
$\Delta\eta \approx \Delta t/a \sim \epsilon f/a\Lambda^2$.
Furthermore, this will continue for a time $\sim H^{-1}$
(since this is the time scale on which $a$ and the amplitude of
$\phi$ are changing), or for a total of $n \sim H^{-1}\Lambda^2/f$
oscillations, from which we can estimate the total growth in $F_\pm$ as
\begin{equation}
  {F_{\pm,f} \over F_{\pm,i}} \sim \exp\left(
  {\epsilon g_{\phi\gamma\gamma}\Lambda^2 \over H} \right)\ .
  \eqnum{4.21}
\end{equation}
Of course, the exponent may contain other factors of order unity, but
this estimate allows us to understand the
dependence of the growth in the magnetic field on the parameters in
the problem.  Note that the analysis is similar if $\Lambda^2 / f \sim
H$, but then $\Delta \eta \approx \Delta t/a$ where $a$ is constant is
only true for time intervals $\Delta t \alt
H^{-1}$.  So using
$\Delta \eta \sim H^{-1}/a$, we see that we get exponential growth
with the same expression in the exponent, but now this exponent is
much smaller.

Since we are interested specifically in long wavelength magnetic
fields, we would also like to know the wavelength at which we get the
most growth by this mechanism.  This follows from our
assumption that the maximum growth occurs at $k \approx
ag_{\phi\gamma\gamma}{\dot
\phi}_{max} \approx ag_{\phi\gamma\gamma}\Lambda^2$, which gives
\begin{equation}
   \lambda \approx {2\pi \over ag_{\phi\gamma\gamma}\Lambda^2}
   \ ,\eqnum{4.22}
\end{equation}
where $\lambda$ is the wavelength today.  More importantly, at the
time the growth occurs the ratio of the wavelength to the
horizon length is given by
\begin{equation}
  {a\lambda \over H^{-1}} = {2\pi aH \over k} \approx {2\pi H
  \over g_{\phi\gamma\gamma}\Lambda^2} \ .\eqnum{4.23}
\end{equation}
But, looking back at (4.21), we see that this ratio is, essentially,
just the inverse of the factor in the exponent, i.e. the larger the
growth in the magnetic field, the smaller the wavelength at which it
occurs!  Further, since an increase in the wavelength by a
factor $\beta$ will result in a decrease in the exponent for the
growth by a factor $\beta^{-1/2}$, it is apparent that significant
growth in the
magnetic field will only occur for wavelengths that are sub-horizon
sized.

We have shown that we cannot have growth in the magnetic field for
super-horizon sized wavelengths when the back effects of the {\bf E}
and {\bf B} fields on $\phi$ are small, but there is still the
possibility that the interaction between the fields and $\phi$ could
allow the magnetic field to be sustained such that it does not
decrease as $a^{-2}$.  However, from (4.7) and (4.8), we see that, if
$\phi$ decays as $a^{-3/2}$ (that is, $\phi$ behaves as a
non-relativistic fluid), then the RHS's of these equations will rapidly
become negligible compared to the individual terms on the LHS's.
We then recover the source-free
Maxwell equations, from which {\bf B} still decreases as
$a^{-2}$.  (The situation is only exacerbated if $\phi$ behaves as a
relativistic fluid.)  In short, there seems to
be no way of sustaining the magnetic field at super-horizon sized
wavelengths.

\section{SUMMARY}

In this paper, we have considered a mechanism for creating a large
scale magnetic field during inflation, proposed by TW, in which the
magnetic field is coupled to a pseudo-Goldstone boson.  We showed
first that the scale required of the magnetic field in order to
explain the galactic magnetic field ($\sim 1$ Mpc) implied that the
growth had to occur at super-horizon sized wavelengths since the
uncoupled equations of motion for a $U(1)$ gauge field imply that the
energy density in the field would simply decay too quickly to be
significant at the end of inflation if there was no enhancement in
the field for wavelengths larger than the horizon.

We then considered the classical evolution of a $U(1)$ gauge field
coupled to a PGB when the back-effects of the field on the PGB were
negligible and showed that such a coupling can, in fact,
produce growth in the field.  However, such growth can occur only at
sub-horizon wavelengths, and thus does not provide a solution to the
above problem.  In the more general case when we allow the
back-effects of the $U(1)$ field to be significant, we still have the
problem that any natural decay of the $\phi$ field ultimately allows
the $U(1)$ field to uncouple from the PGB leaving  us once again with
a free $U(1)$ field.
Hence, it seems to be impossible to create a significant magnetic
field by simply coupling the magnetic field to a PGB during inflation.

\nonum
\section{ACKNOWLEDGEMENTS}

We thank Bill Press for helpful discussions.  Support
for this work was provided by NASA under grants NAGW-931 and
NGT-50850, and by the National Science Foundation.

\end{document}